\documentclass[12pt]{article}

\def\tr{{\rm tr}}

\begin{document}

\title{Late-time oscillatory behaviour for self-gravitating scalar fields}

\author{Alan D. Rendall\\
Max-Planck-Institut f\"ur Gravitationsphysik\\Albert-Einstein-Institut
\\ Am M\"uhlenberg 1\\
14476 Golm, Germany}

\date{}

\maketitle

\begin{abstract}
This paper investigates the late-time behaviour of certain cosmological 
models where oscillations play an essential role. Rigorous results are 
proved on the asymptotics of homogeneous and isotropic 
spacetimes with a linear massive scalar field as source. Various
generalizations are obtained for nonlinear massive scalar fields,
$k$-essence models and $f(R)$ gravity. The effect of adding ordinary matter
is discussed as is the case of nonlinear scalar fields whose potential 
has a degenerate zero.
\end{abstract}

\section{Introduction}

Scalar fields are important in cosmology as a mechanism for producing models 
with accelerated expansion, both in the very early universe (inflation) and
in more recent epochs. The mathematical properties of spatially 
homogeneous solutions of the Einstein equations coupled to different types
of scalar field and ordinary matter have been studied in a number of papers
\cite{kitada}, \cite{lee}, \cite{rendall04}, \cite{rendall05}, 
\cite{rendall06}, \cite{bieli} in the case that there is accelerated expansion 
in the whole future of some late time. A corresponding analysis of situations 
where the accelerated expansion is only temporary, being succeeded by 
decelerated expansion, has not yet been carried out. One reason for this is 
that in contrast to what is found for models with continuing accelerated 
expansion the behaviour of the simplest models, which are homogeneous, 
isotropic and spatially flat, does not carry over to more general 
homogeneous models. In this paper the very simplest case of this type, the 
massive linear scalar field, is analysed. Then a number of other cases showing
similar late-time behaviour are discussed.

Consider solutions of the Einstein equations coupled to a massive scalar 
field which are homogeneous, isotropic and spatially flat. The standard
picture of the dynamics of these solutions is as follows \cite{belinskii1}.
(See also the earlier papers \cite{starobinskii1} and \cite{starobinskii2}.)
The solutions exhibit accelerated expansion at intermediate times and
eventually enter a phase where they are on average decelerated and resemble 
dust solutions. The main aim of this paper is to prove rigorous results
about the late-time behaviour. Formulae for this can be found in the 
literature. (See for instance \cite{mukhanov}.) Here a proof will be presented 
that these formulae do provide asymptotic expansions for all solutions in a 
precise sense. A curious fact, which does not seem to be mentioned in the
literature, is that it is a consequence of the asymptotic expressions that 
each solution has infinitely many phases of accelerated expansion. The early
and intermediate time behaviour will not be discussed further here but note
that some rigorous results on this were obtained in \cite{rendall02}. 

A key reason why the massive scalar field is more difficult to handle
than types of scalar field which have previously been treated mathematically 
is the following. When deriving asymptotic expansions in this case it is
repeatedly necessary to estimate integrals with upper limit infinity 
which are convergent but not absolutely convergent. The convergence of
the integral, which is required for the proof, results from the 
cancellation of positive and negative contributions. In other words, 
there are oscillations in the solution which continue to have a
significant effect for all times.

The paper is organized as follows. In Section 2 asymptotic expansions are 
proved for the basic example, the massive scalar field. This is generalized 
to a large class of nonlinear massive scalar fields in Section 3. It is 
shown that a number of features of the asymptotics are unchanged and the 
expansion is derived up to the point where the first modification occurs.
It is discussed briefly how these results can be applied to $f(R)$ theories
of gravity. Analogous results for certain $k$-essence models are also 
derived. In all of this the only matter present is the scalar field. The
question of the incorporation of ordinary matter such as a perfect fluid
is discussed in Section 4. In Section 5 it is investigated what happens 
for a scalar field whose potential has a degenerate minimum where it 
vanishes. Conclusions and possible extensions of the results obtained
are the subject of the last section.

\section{The massive scalar field} 

This section is concerned with a massive linear scalar field $\phi$ 
minimally coupled to the Einstein equations under the assumptions of a 
spatially flat Friedmann model and the absence of any other matter fields.
To simplify the computations it is assumed that the scalar field has unit
mass. The scale factor is denoted by $a$ and the Hubble parameter is given 
by $H=\dot a/a$. The equation of motion of the scalar field is
$\ddot\phi+3H\dot\phi+\phi=0$ and the Hamiltonian constraint is
$H^2=\frac{4\pi}3 (\dot\phi^2+\phi^2)$. The evolution equation for $H$
is $\dot H=-4\pi\dot\phi^2$. If $H$ is zero at any time $t_1$ then the 
initial data for the scalar field at time $t_1$ vanishes. Thus the scalar 
field vanishes at all times as does $H$. This is just flat space and so from 
now on it will be assumed that $H$ never vanishes. It must have a constant 
sign and this sign is chosen to be positive corresponding to an expanding 
model. The equation of motion of $\phi$ can be rewritten as
\begin{equation}\label{eom}
\ddot\phi+\sqrt{12\pi}(\dot\phi^2+\phi^2)^{1/2}\dot\phi+\phi=0.
\end{equation}

The function $H$ cannot increase and so it must tend to a limit $H_0$ as 
$t\to\infty$. It will now be shown that $H_0$ vanishes. Assume that $H_0>0$ 
with the aim
of deriving a contradiction. The functions $\phi$ and $\dot\phi$ are both
bounded as is $H$. From the equation of motion it follows that the same
is true of $\ddot\phi$. Differentiating the equation repeatedly with repect 
to time shows that higher derivatives of $\phi$ are also bounded. Let 
$\{t_n\}$ be a sequence tending to infinity and let $\phi_n(t)=\phi(t+t_n)$ 
and $H_n(t)=H(t+t_n)$. Using the bounds already listed, the Arzela-Ascoli 
theorem \cite{rudin} can be applied. This implies that, after passing to a 
subsequence, $\phi_n$ converges uniformly on compact sets to a limit 
$\phi_\infty$. 
Moreover the first and second derivatives of $\phi_n$ converge to the 
corresponding derivatives of $\phi_\infty$. The sequence $H_n$ converges 
uniformly to the constant value $H_0$. With this information  it is possible 
to pass to the limit in the equation (\ref{eom}) to obtain
\begin{equation}\label{eominfty}
\ddot\phi_\infty+3H_0\dot\phi_\infty+\phi_\infty=0
\end{equation}
From the Hamiltonian constraint it follows that 
$H^2_0=\frac{4\pi}3 (\dot\phi_\infty^2+\phi_\infty^2)$. Differentiating this
relation with respect to $t$ and substituting (\ref{eominfty}) into the result
shows that  $\dot\phi_\infty=0$. Then using (\ref{eominfty}) again shows that
$\phi_\infty=0$ and this leads to a contradiction to the assumption that 
$H_0\ne 0$. Thus in fact $H(t)\to 0$ as $t\to\infty$.

Next more detailed asymptotics will be obtained. The pair 
$$(\phi/(\dot\phi^2+\phi^2)^{1/2}, \dot\phi/(\dot\phi^2+\phi^2)^{1/2})$$
defines
a function of $t$ with values in the unit circle. Thus it is possible to 
define a real-valued function $\theta (t)$ whose projection to the circle
under identification modulo $2\pi$ is the given function with values in
the circle. The function $\theta$ is unique up to shifts of its argument 
by integer multiples of $2\pi$. Let $r=(\dot\phi^2+\phi^2)^{1/2}$. Then the 
following equations can be derived:
\begin{eqnarray}
\dot r&=&-\sqrt{12\pi}r^2\sin^2\theta           \\
\dot\theta&=&-1-\sqrt{12\pi}r\sin\theta\cos\theta
\end{eqnarray}
Given $\epsilon>0$ there exists a time $t_1$ such that 
$\sqrt{12\pi}r\le\epsilon$ for $t\ge t_1$. It follows that on that interval
$|\dot\theta+1|\le\epsilon$. This means that
\begin{equation}
\theta_1-(1+\epsilon)(t-t_1)\le\theta (t)\le\theta_1-(1-\epsilon)(t-t_1)
\end{equation}
where $\theta_1=\theta(t_1)$. The evolution equation for $r$ can be rewritten
in the form
\begin{equation}\label{rinverse}
r^{-1}(t)=r^{-1}(t_1)+\sqrt{3\pi}\left[(t-t_1)+\int_{\theta (t)}^{\theta_1}
\cos 2\theta' \dot\theta^{-1}(\theta')d\theta'\right]
\end{equation}
Suppose that $\epsilon<1$. Let $k$ be the largest integer such that 
$\theta_1-k\pi\ge \theta(t)$. Then
\begin{equation}
\left|\int_{\theta (t)}^{\theta_1-k\pi}
\cos 2\theta' \dot\theta^{-1}(\theta')d\theta'\right|
\le \pi (1-\epsilon)^{-1}
\end{equation}
On the other hand 
\begin{eqnarray}
&&\left|\int_{\theta_1 -k\pi}^{\theta_1}
\cos 2\theta' \dot\theta^{-1}(\theta')d\theta'\right| \nonumber\\
=&&\left|\int_{\theta_1-k\pi}^{\theta_1}
\cos 2\theta' [\dot\theta^{-1}(\theta')+1]d\theta'\right| \nonumber\\
\le &&k\pi \epsilon(1-\epsilon)^{-1} \nonumber\\
\le &&\epsilon(1+\epsilon)(1-\epsilon)^{-1}(t-t_1)
\end{eqnarray}
Thus the integral on the right hand side of (\ref{rinverse}) is bounded
in modulus by $(1-\epsilon)^{-1}(\pi+\epsilon(1+\epsilon)(t-t_1))$. Putting 
this information into (\ref{rinverse}) shows that
\begin{equation}
r^{-1}(t)\ge r^{-1}(t_1)+\sqrt{3\pi}\left\{-(1-\epsilon)^{-1}\pi+
[1-\epsilon(1+\epsilon)(1-\epsilon)^{-1}](t-t_1)\right\}
\end{equation}
As a consequence $r^{-1}$ grows linearly with $t$ and 
$r(t)=O(t^{-1})$. The evolution equation for $\theta$ then implies that
$\dot\theta+1=O(t^{-1})$ and that
\begin{equation}\label{theta1}
\theta(t)=-t+O(\log t)
\end{equation}
Putting the improved estimate for $\dot\theta$ back into (\ref{rinverse})
shows that 
\begin{equation}\label{r1}
r(t)=\frac1{\sqrt{3\pi}t}+O(t^{-2}\log t).
\end{equation}
Translating this to the original variables it can be seen that the leading 
order contribution to $H$ is $2/3t$ and that the scale factor behaves 
asymptotically like $t^{2/3}$. 
Thus the late-time dynamics is similar to that of a model with
dust. To go further it is necessary to obtain an improved estimate for 
$\theta$. Note that 
\begin{equation}\label{improved}
\frac{d}{dt} (r\cos 2\theta)=\dot r\cos 2\theta-2r\sin 2\theta\dot\theta
=2r\sin 2\theta+O(t^{-2})
\end{equation}
Integrating the equation for $\dot\theta$ and using the above relation gives
\begin{equation}\label{theta2}
\theta=-t+C+O(t^{-1}).
\end{equation}
for a constant $C$. This constant can be eliminated by choosing an appropriate
origin for $t$. Then $\theta=-t+O(t^{-1})$.

Consider now the asymptotic behaviour of the scalar field. 
\begin{eqnarray}
\cos (\theta(t))&=&\cos t\cos (\theta(t)+t)+\sin t\sin(\theta(t)+t)\nonumber
\\
&=& \cos t+O(t^{-1})
\end{eqnarray}
Hence
\begin{equation}\label{phi1}
\phi(t)=\frac{\cos t}{\sqrt {3\pi}t}+O(t^{-2}\log t)
\end{equation}
Similarly
\begin{equation}\label{phi2}
\dot\phi(t)=\frac{\sin t}{\sqrt {3\pi}t}+O(t^{-2}\log t)
\end{equation}

Putting the expressions for $\sin\theta$ and $\cos\theta$ into the equation
for $\dot\theta$ gives:
\begin{equation}\label{thetadot}
\dot\theta=-1-t^{-1}\sin 2t+O(t^{-2}\log t)
\end{equation}
This can now be substituted into (\ref{improved}) to get
\begin{equation}\label{rcos}
\frac{d}{dt} (r\cos 2\theta)=2r\sin 2\theta-\frac2{\sqrt{3\pi}}t^{-2}
\sin^2 t\cos 2t+\frac2{\sqrt{3\pi}}t^{-2}\sin^2 2t+O(t^{-3}\log t)
\end{equation}
The following lemma is useful for estimating the integral of oscillatory
quantities. It is proved by integrating by parts twice.

\noindent
{\bf Lemma 1} Let $F$ be a smooth periodic function with mean zero, $f=F'$ and 
$k$ a positive real number. Then 
\begin{equation}
\int_1^T f(t)t^{-k} dt=F(T)T^{-k}+C+O(T^{-k-1})
\end{equation}
for a constant $C$.

Integrating (\ref{rcos}), applying Lemma 1 and noting that due to the 
information already obtained the integration constant must vanish, it
follows that
\begin{equation}
\theta(t)=-t-\frac{3+2\cos 2t}{4t}+O(t^{-2}\log t)
\end{equation}
Putting the information now available into the equation for 
the time derivative of $r^{-1}$ gives
\begin{equation} 
\frac{d}{dt} (r^{-1})=\sqrt{3\pi}\left(1-\cos 2t+\sin 2t
\left(\frac{3+2\cos 2t}{2t}\right)+O(t^{-2}\log t)\right)
\end{equation}
and after integration
\begin{equation}
r^{-1}=\sqrt{3\pi}\left[(t-t_1)-\frac{\sin 2t}{2}\right]+O(t^{-1}\log t)
\end{equation}
for a constant $t_1$. It follows that 
\begin{equation}
H=\frac{2}{3(t-t_1)}\left[1+\frac{\sin 2t}{2t}
+O(t^{-2}\log t)\right]
\end{equation}
Note that the constant $t_1$ in the leading term in the expression for $H$
could be got rid of by a translation in $t$ at the expense of introducing a 
similar constant elsewhere. The expressions for the scalar field can now be 
improved to give 
\begin{equation}
\phi(t)=\frac1{\sqrt{3\pi}(t-t_1)}\left[\cos t-\sin t\left(
\frac{3+2\cos 2t}{4t}\right)\right]\left[1+\frac{\sin 2t}{2t}\right]
+O(t^{-3}\log t)
\end{equation}
and 
\begin{equation}
\dot\phi(t)=\frac1{\sqrt{3\pi}(t-t_1)}\left[-\sin t-\cos t\left(
\frac{3+2\cos 2t}{4t}\right)\right]\left[1+\frac{\sin 2t}{2t}\right]
+O(t^{-3}\log t)
\end{equation}
The first of these may be compared with equation (5.45) on p. 240 of
\cite{mukhanov}.

Since the scale factor grows slower than linearly the 
expansion is on average decelerated. On the other hand 
$\ddot a=\frac{4\pi}3 (\phi^2-2\dot\phi^2)$ and so there are infinitely
many intervals on which $\ddot a>0$.

\section{Nonlinear scalar fields}

The aim of this section is to investigate to what extent the results for the
massive linear scalar field can be extended to the case of a potential of
the form $V(\phi)=\phi^2/2+W(\phi)$ where $W$ is smooth and 
$W(\phi)=O(\phi^3)$. The equation of motion for the scalar field is
\begin{equation}
\ddot\phi+\sqrt{12\pi}(\dot\phi^2+\phi^2+2W(\phi))^{1/2}\dot\phi
+\phi+W'(\phi)=0
\end{equation}
The evolution equation for $H$ is unchanged. It can be concluded that $H$
tends to a limit $H_0$ as $t\to\infty$. To constrain the value of $H_0$
a compactness argument can be used as before, leading to the limiting 
equation
\begin{equation}
\ddot\phi_\infty+3H_0\dot\phi_\infty+\phi_\infty+W'(\phi_\infty)=0
\end{equation}
with $H^2_0=\frac{4\pi}{3}(\dot\phi_\infty^2+\phi_\infty^2+2W(\phi_\infty))$. 
It follows that $\dot\phi_\infty=0$. Assume now that $V$ has no critical 
points other than the origin in an interval $[-\phi_1,\phi_1]$ and that the 
initial value of $H$ is smaller than $\sqrt{\frac{4\pi}{3}}\phi_1$ so that 
$\phi$ remains in that
interval. Then it follows that $\phi_\infty=0$ and this gives a contradiction. 
Thus in fact $H_0=0$. As in the linear case polar coordinates $(r,\theta)$ 
can be introduced in the $(\phi,\dot\phi)$-plane. In these variables the 
equations read
\begin{eqnarray}
\dot r&=&-\sqrt{12\pi}r^2\sin^2\theta (1+2r^{-2}W(r\cos\theta))^{1/2}
-\sin\theta W'(r\cos\theta)           \\
\dot\theta&=&-1-\sqrt{12\pi}r\sin\theta\cos\theta(1
+2r^{-2}W(r\cos\theta))^{1/2}
\nonumber                   \\
&&-r^{-1}W'(r\cos\theta)\cos\theta
\end{eqnarray}
Note that $\dot\theta+1=O(r)$ so that it can be concluded as in the case 
$W=0$ that $\theta$ grows linearly with $t$. The equation for $\dot r$ can 
be rearranged to give
\begin{equation}
\frac{d}{dt}(r^{-1})=\sqrt{12\pi}\sin^2\theta (1+2r^{-2}W(r\cos\theta))^{1/2}
+\sin\theta r^{-2}W'(r\cos\theta)
\end{equation}
The last term in this equation looks worrying but note that up to a 
remainder of order $r$ it is equal to 
$\frac12 W'''(0)\cos^2\theta\sin\theta$. The last expression has mean zero 
and so its integral up to time up $t$ is a bounded function of $t$, as can 
be seen by changing the variable of integration from $t$ to $\theta$. It can 
be concluded as in the linear case that $r^{-1}$ grows linearly with $t$
and that $r(t)=O(t^{-1})$. Furthermore $\dot\theta+1=O(t^{-1})$
and (\ref{theta1}) and (\ref{r1}) hold. Thus $H(t)$ has the same leading 
order behaviour as for dust in this case too.

Next further information on the asymptotics of $\theta$ will be obtained,
reaching the point where the first correction coming from $W$ occurs. Note 
first that $(1+2r^{-2}W(r\cos\theta))^{1/2}=1+O(r)$ so that the correction 
coming from $W$ in the second term on the right hand side of the equation
for $\dot\theta$ is $O(t^{-2})$. The last expression in (\ref{improved}) 
can be used to replace $d/dt(r\cos 2\theta)$ as in the case $W=0$ and this 
allows the integral of $r\sin 2\theta$ to be treated. To 
handle the last term in the equation for $\dot\theta$ note that
\begin{equation}
r^{-1}W'(r\cos\theta)\cos\theta=\frac1{6\sqrt{3}} t^{-1}W'''(0)
\frac{d}{d\theta}\left(3\cos\theta-\sin^3\theta\right)+O(t^{-2}\log t)
\end{equation}
Furthermore
\begin{equation}
\frac{d}{d\theta} (3\cos\theta-\sin^3\theta)=\frac{d}{dt}(3\cos\theta
-\sin^3\theta)+O(t^{-1})
\end{equation}
Using Lemma 1 this implies that $\theta=-t+O(t^{-1}\log t)$. The relations 
(\ref{phi1}) and (\ref{phi2}) follow. The first place where $W$ makes a 
difference in the asymptotic expansions is in the analogue of 
(\ref{thetadot}). It is given by
\begin{equation}
\dot\theta=-1-t^{-1}\sin 2t-\frac1{2\sqrt{3}} t^{-1}W'''(0)\cos^3 t
+O(t^{-2}\log t)
\end{equation} 

The results obtained above in the case $m=1$ can be generalized to any 
positive $m$. To get the asymptotic formulae in the general case it suffices
to replace $t$ by $mt$ everywhere. Theorems are obtained for the potentials
$\frac12 m^2\phi^2+\lambda\phi^4$ and $\frac14\lambda (\phi^2-\phi_0^2)^2$
considered in \cite{belinskii1}. In the first case the late-time asymptotics
are proved for all solutions while in the second they are proved for all
solutions starting sufficiently close to one of the two local minima of the
potential. 

These results can also be applied to obtain information
about the $f(R)$ theories of gravity which are equivalent to the 
Einstein equations coupled to a nonlinear scalar field via a conformal 
transformation. For a discussion of this see section 5.6 of \cite{mukhanov}.
For example consider the theory where the Einstein-Hilbert 
Lagrangian is replaced by $R+\alpha R^2$ for a negative constant $\alpha$. The
potential of the corresponding scalar field is non-negative and has a
unique minimum at zero. Its second derivative at that point is $-8\pi/3\alpha$.
Hence the theory just developed can be applied to this case. The
interval $[-\phi_1,\phi_1]$ and a bound for the value of $H$ at the initial
time must be chosen appropriately. Similar arguments work for more general
choices of $f(R)$. Assume that $f(0)=0$, $\frac{df}{dR}(0)=1$ and 
$\frac{d^2 f}{dR^2}(0)<0$. Then the corresponding scalar field has the 
properties which allow the results of this section to be applied to it.
Since $\phi\to 0$ as $t\to\infty$ the conformal factor tends to one as
$t\to\infty$ and the physical metric has the same leading order
asymptotics as the conformally rescaled one.

It may be remarked in passing that the results of \cite{rendall04} also have
applications to determining the late-time behaviour of $f(R)$ models
with continuing acceleration. If there is a positive constant $V_0$ such that 
$f(-2V_0)=-V_0$ and $f'(-2V_0)=1$ then the corresponding potential has 
a minimum at zero where it takes the value $V_0$. Provided $f''(-2V_0)$
lies in the interval $(-\frac{1}{2V_0},0)$ then the second derivative of the
potential at zero is positive and the results of \cite{rendall04} can be
applied. Note that the case $f(R)=-R^2$ discussed in \cite{starobinsky}
(without symmetry assumptions on the solutions) is a borderline one since
$f''(-2V_0)=-\frac1{2V_0}$. In that case the potential is constant. 

Another type of generalization is to $k$-essence where the Lagrangian
$X-V$ of an ordinary nonlinear scalar field, with
$X=-\frac12 \nabla_\alpha\phi\nabla^\alpha\phi$, is replaced by a 
more general function $L(\phi,X)$. The evolution equation for 
$H$ with $H^2=\frac{8\pi}3(2X\partial L/\partial X-L)$ becomes
$\dot H=-8\pi X\partial L/\partial X$. Note that in a spatially homogeneous 
solution $X$ is non-negative. This Lagrangian will now be specialized to the 
case $L(\phi,X)=X-\frac12\phi^2-W(\phi,X)$ where $W$ vanishes at the origin 
up to a remainder of third order. Now 
\begin{equation}
2X\partial L/\partial X-L=X+\frac12\phi^2-2X\partial W/\partial X+W.
\end{equation}
It follows that there is a constant $C>0$ such that 
\begin{equation}
C^{-1}(\phi^2+X)\le 2X\partial L/\partial X-L\le C(\phi^2+X)
\end{equation}
for $\phi$ and $X$ small. This means in particular that when $\phi$ and $X$
are small $H$ can only be zero when both $\phi$ and $X$ are zero. As 
a consequence the initial data for $\phi$ vanishes so that $\phi$ vanishes 
everywhere. This case will be excluded from consideration and so it can be
assumed as in the cases studied previously that $H>0$. Under the given 
assumptions on $W$ it follows that $H$ is non-increasing. Thus if it
starts small it remains small and tends to a limit $H_0$ as $t\to\infty$.
The equation of motion of $\phi$ is
\begin{equation}
\left(\frac{\partial L}{\partial X}+2X\frac{\partial^2 L}{\partial X^2}\right)
\ddot\phi+\frac{\partial L}{\partial X}(3H\dot\phi)+\frac{\partial^2 L}
{\partial\phi\partial X}\dot\phi^2-\frac{\partial L}{\partial\phi}=0
\end{equation}
In order to repeat the compactness argument which has been used in other
cases all that needs to be ensured is that the coefficient of 
$\ddot\phi$ in this equation remains bounded away from zero. If the initial
value of $H$ is small enough this is the case. Passing to the limit in the 
evolution equation for $H$ it follows that $\dot\phi_\infty=0$. The equation
of motion for $\phi_\infty$ implies that 
$\frac{\partial L}{\partial\phi} (\phi_\infty,0)=0$. For $\phi_\infty$ 
sufficiently small this means that $\phi_\infty=0$. Hence $H_0=0$. 

Passing to polar coordinates in the $(\phi,\dot\phi)$ plane the equation of
motion becomes:
\begin{eqnarray}
&&\dot r=\sin\theta\left[1-\frac{\partial W}{\partial X}-2X
\frac{\partial^2 W}{\partial X^2}\right]^{-1}
\left[-\left(1-\frac{\partial W}{\partial X}\right)(3H\dot\phi)
+\frac{\partial^2 W}{\partial\phi\partial X}\dot\phi^2
-\frac{\partial W}{\partial\phi}\right]\nonumber                    \\
&&-r\cos\theta\sin\theta\left[\frac{\partial W/\partial X
+2X\partial^2W/\partial X^2}{1-\partial W/\partial X
-2X\partial^2W/\partial X^2}\right]                                 \\
&&\dot\theta=-1-\cos^2\theta\left[\frac{\partial W/\partial X
+2X\partial^2 W/\partial X^2}{1-\partial W/\partial X
-2X\partial^2 W/\partial X^2}\right]                                
+\cos\theta\left[1-\frac{\partial W}{\partial X}
-2X\frac{\partial^2 W}{\partial X^2}\right]^{-1}\nonumber           \\
&&\times\left[-\left(1-\frac{\partial W}{\partial X}\right)(3H\sin\theta)
+\frac{\partial^2 W}{\partial\phi\partial X}r\sin^2\theta
-r^{-1}\frac{\partial W}{\partial\phi}\right]
\end{eqnarray}
The expression $\partial W/\partial X+2X\partial^2 W/\partial X^2$,
when evaluated at $(r\cos\theta,r^2\sin^2\theta/2)$, is $O(r^2)$. From this
it easily follows that $\dot\theta+1$ is $O(r)$ and so $\theta$ grows 
linearly with $t$. The last term in the equation for $\dot r$ is $O(r^3)$, 
as is the expression $\dot\phi^2\partial^2 W/\partial\phi\partial X$. Thus 
\begin{equation}
\frac{d}{dt}(r^{-1})=\sqrt{12\pi}\sin^2\theta+\sin\theta r^{-2}
\frac{\partial W}{\partial\phi}+O(r).
\end{equation}
The second term on the right hand side of this equation can be written in
the form 
$\frac12\frac{\partial^3 W}{\partial\phi^3}(0,0)\cos^2\theta\sin\theta
+O(r^{-1})$.
It can be concluded as in the case of the ordinary nonlinear scalar field
that $r=O(t^{-1})$ and that $H$ has the same leading order behaviour as in 
the case of dust.

\section{Inclusion of matter}\label{matter}

The previous sections were concerned with the Einstein equations coupled
to a scalar field (possibly nonlinear) without including other fields
describing ordinary matter. In this section matter satisfying the 
dominant and strong energy conditions will be added. The energy-momentum
tensor is the sum of that of a scalar field and that of the other matter.
There is no direct coupling between the scalar field and the ordinary
matter - they interact only indirectly via their coupling to the
gravitational field. The assumption of a spatially flat Friedmann model
is maintained. The equation of motion of the scalar field, before 
substituting for $H$ using the Hamiltonian constraint, remains unchanged.
The Hamiltonian constraint reads
\begin{equation}
H^2=\frac{4\pi}{3}(\dot\phi^2+\phi^2+2V+2\rho_M)
\end{equation}
where where $\rho_M$ is the energy density of the ordinary matter. If $H$ is 
zero at any time $t_1$ the initial data for the scalar field vanishes as a
consequence of the weak energy condition. Then the scalar field vanishes at
all times. Since the subject of interest here is the effect of a scalar field 
that case will be excluded. It is assumed from now on that $H$ is positive.
The evolution equation for $H$ is 
\begin{equation}\label{hdotmatter}
\dot H=-4\pi\left[\dot\phi^2+\rho_M+\frac13\tr S_M\right]
\end{equation}
where $\tr S_M$ is the trace of the spatial projection of the energy-momentum
tensor of the ordinary matter. The dominant energy condition implies that
the right hand side of (\ref{hdotmatter}) is non-positive so that $H(t)$
is non-increasing. 

Assuming that the solution of the coupled Einstein-scalar-matter
equations exists globally in the future it can be concluded that $H(t)$ tends 
to a limit $H_0$ as $t\to\infty$. The quantities $\phi$, $\dot\phi$ and $H$
are all bounded. Hence $\ddot\phi$ is bounded. The energy-momentum tensor
of the ordinary matter is divergence-free and it follows that
\begin{equation}\label{continuity}
d\rho_M/dt+3H\left(\rho_M+\frac13\tr S_M\right)=0
\end{equation} 
Thus, by the dominant energy condition, $\rho_M$ is non-increasing and 
converges to some limit $\rho_\infty\ge 0$. Also 
$\tr S_M\le 3\rho_M$ and it follows that $\dot H$ is bounded. Differentiating
the equation of motion for the scalar field shows that the third time 
derivative of $\phi$ is bounded. Using these facts a compactness argument
can be carried out as in the case without ordinary matter. Under the same
conditions it can be concluded that $\phi$ and $\dot\phi$ tend to zero as
$t\to\infty$. Moreover it follows from (\ref{hdotmatter}) that 
$\rho_M+\frac13\tr S_M\to 0$. Hence $\tr S_M\to -3\rho_\infty$ as 
$t\to\infty$. But then $\rho_M+\tr S_M\to -2\rho_\infty$. If $\rho_\infty$
were non-zero then this would contradict the strong energy condition. Hence
in fact $\rho_\infty=0$, $H_0=0$ and $\rho_M(t)\to 0$ as $t\to\infty$.

Consider for simplicity the linear case $W=0$. When the 
equation of motion for $\phi$ is written in polar coordinates the
resulting equations are
\begin{eqnarray}
&&\dot r=-\sqrt{12\pi}(r^2+2\rho_M)^{1/2}r\sin^2\theta      \\
&&\dot\theta=-1-\sqrt{12\pi}(r^2+2\rho_M)^{1/2}
\sin\theta\cos\theta
\end{eqnarray}
Since $H$ tends to zero as $t\to\infty$ the argument that $\theta$ grows
linearly with $t$ also works in the presence of ordinary matter. It is
also true that $\dot\theta+1$ remains bounded away from zero. Analysis 
of the evolution equation for $r^{-1}$ shows that $r^{-1}$ must grow at
least linearly and that $r=O(t^{-1})$. Due to its sign the contribution of 
the energy density of ordinary matter can only increase this rate of decay.
Unfortunately this does not immediately give a decay rate for $H$. It would 
seem that to go further it would be necessary to know which of the terms 
$r^2$ or $\rho_M$ dominates at late times. In the case of a perfect fluid
with linear equation of state $p_M=(\gamma-1)\rho_M$ heuristic 
considerations indicate that if $\gamma>1$ it is self-consistent to
require that the fluid has a negligible effect at late times while for
$2/3<\gamma<1$ (a model which satisfies the strong and dominant energy 
conditions) this is not consistent. Rigorous results on the asymptotics
in these cases are not available.

\section{Degenerate minima}

Suppose now that the linear scalar field is replaced by a nonlinear one with
$V(\phi)=\phi^{2n}/2n$ and no additional matter is included. Then the 
equation for $\dot H$ is unchanged and the equation of motion is
\begin{equation}\label{eomn}
\ddot\phi+\sqrt{12\pi}(\dot\phi^2+\phi^{2n}/n)^{1/2}\dot\phi+\phi^{2n-1}=0
\end{equation}
As in the case $n=1$ the function $H$ is non-increasing and tends to a limit
$H_0$. The argument that $H_0=0$ goes through without significant change so
that $\phi$ and $\dot\phi$ tend to zero as $t\to\infty$. For simplicity only 
the special case $n=2$ will be considered in the rest of this paragraph. 
Passing to polar coordinates leads to the system
\begin{eqnarray}
&&\dot r=r\sin\theta\cos\theta (1-r^2\cos^2\theta)-3Hr\sin^2\theta  \\
&&\dot\theta=-\sin^2\theta-r^2\cos^4\theta-3H\sin\theta\cos\theta          
\end{eqnarray}
where 
\begin{equation}
H=\sqrt{\frac{4\pi}3}r\left(\sin^2\theta+\frac12 r^2\cos^4\theta\right)^{1/2}
\end{equation}
For the solutions being considered here $r$ never vanishes at any time.
It follows that whenever $\theta=k\pi$ or $(k+1/2)\pi$ for $k$ an integer
$\dot\theta<0$. Hence there are two mutually exclusive possibilities. Either
$\theta(t)\to -\infty$ as $t\to\infty$ or $\theta(t)$ is eventually
trapped between $\theta_1$ and $\theta_1+\pi/2$ where $2\theta_1/\pi$ is an
integer. It will now be shown that the second of these cannot occur. To
do this it is helpful to distinguish between the case where $\theta_1/\pi$ is
an integer and that where $\theta_1/\pi+1/2$ is an integer. In the first case 
$\theta$ is monotone decreasing at late times and $\theta\to\theta_1$ as
$t\to\infty$. For $r$ small enough
\begin{equation}
\dot r\ge r\sin\theta\left(\frac12\cos\theta-3H\sin\theta\right)
\end{equation}
This implies that $\dot r$ is eventually positive, contradicting the fact
that $H\to 0$ as $t\to\infty$. Thus this case is ruled out and it can 
be assumed that $\theta_1/\pi+1/2$ is an integer. In that case $r$ is
eventually monotone decreasing and it can be concluded that $r=o(1)$.
For any interval of the form $[\theta_1,\theta_2]$ with
$\theta_2<\theta_1+\pi/2$ it is true that at sufficiently late times any 
solution for which $\theta$ lies in this interval satisfies a uniform
negative upper bound on $\dot\theta$. Thus $\theta\to\theta_1+\pi/2$ as 
$t\to\infty$ and $H=o(r)$. Thus for any $\epsilon>0$ there is a time $t_1$ 
such that for $t>t_1$ 
\begin{eqnarray}
&&-\sin^2\theta-r^2\cos^4\theta-3H\sin\theta\cos\theta  \nonumber      \\
&&\le -\sin^2\theta-r^2\cos^4\theta+\epsilon r|\sin\theta||\cos\theta|\nonumber
 \\
&&\le -\sin^2\theta\left(1-\frac\epsilon{2}\right)-r^2\cos^2\theta\left(
\cos^2\theta-\frac\epsilon{2}\right)\nonumber                          \\
&&<0
\end{eqnarray}    
Using the evolution equation for $\theta$ this gives a contradiction. Thus it 
can be concluded that in fact $\theta (t)\to -\infty$ as $t\to\infty$. It
follows that there are infinitely many oscillations of the scalar field in 
this case too.

There is a remark on p. 242 of \cite{mukhanov} that in the case of a nonlinear
scalar field with potential $V(\phi)=\phi^{2n}/2n$ the solution should be 
approximated in some averaged sense by a perfect fluid with linear equation
of state. A statement will now be proved which is a concrete realization of
this idea. Multiplying the equation of motion for a nonlinear scalar field
and rearranging gives the identity
\begin{equation}
\phi V'(\phi)=\dot\phi^2-6\pi\phi^2\dot\phi^2-\frac{d}{dt}\left(\phi\dot\phi
+\frac32 H\phi^2\right)
\end{equation}
In deriving this the equation for $\dot H$ has been used. The information 
already available concerning the solution justifies integrating this relation
from some time $t$ to infinity to get
\begin{equation}\label{phi4identity}
\int_t^\infty \phi V'(\phi) ds=\int_t^\infty \dot\phi^2-6\pi\dot\phi^2\phi^2
ds-\phi(t)\dot\phi(t)
\end{equation}
In the special case of the power-law potential $\phi V'(\phi)=2nV(\phi)$.
Integrating the evolution equation for $H$ and using the fact that $H(t)$
tends to zero as $t\to\infty$ shows that
\begin{equation}
\int_t^\infty\dot\phi^2 ds=\frac{1}{4\pi}H(t)
\end{equation}
Using this in (\ref{phi4identity}) shows that 
\begin{equation}
\int_t^\infty\phi V'(\phi) ds=\left(\int_t^\infty\dot\phi^2 ds\right) 
(1+o(1)),\ \ \ t\to\infty
\end{equation}
It follows that
\begin{equation}
\frac{\int_t^\infty p(s)ds}{\int_t^\infty \rho(s) ds}
=\left(\frac{n-1}{n+1}\right)(1+o(1))
\end{equation}
This corresponds to dust in the case $n=1$ and to radiation in the case
$n=2$.

\section{Conclusions}

In this paper rigorous asymptotic expansions for the late-time asymptotics 
of spatially flat homogeneous and isotropic solutions of the Einstein
equations coupled to a linear massive scalar field are proved. It is shown 
what similarities and differences there are when the quadratic potential of 
this model is modified by higher order corrections. The latter results
apply directly to give information about the dynamics in certain $f(R)$ 
theories. Basic asymptotics are also obtained for $k$-essence models where 
the leading order terms in the Lagrangian near the origin agree with those 
of a linear massive scalar field.

When ordinary matter such as a perfect fluid is added to the model some
statements about the asymptotic behaviour are proved but at a certain stage
in the expansion a competition arises between the scalar field and the other 
matter and the outcome of this is not settled rigorously here. In the case
of a potential such as $V(\phi)=\phi^4/4$ with a degenerate minimum it is
shown that the scalar field is oscillatory in the sense that it has infinitely 
many zeroes. A relation to the radiation fluid is derived.

As has already been indicated, the asymptotics derived in section 2 can
be destroyed by the introduction of spatial curvature. When homogeneous
solutions with non-vanishing spatial curvature are considered the equation
of motion for $\phi$ remains the same but the expression for $H$ given by
the Hamiltonian constraint and the evolution equation for $H$ pick up extra 
contributions involving shear and spatial curvature. 
For Bianchi types I-VIII where $R\le 0$
the extra contributions to $H$ and $\dot H$ are positive and negative 
respectively. As in the isotropic and spatially flat case it can be 
assumed without loss of generality that $H>0$ while $H$ is non-increasing
and tends to a limit $H_0\ge 0$. A compactness argument as in previous 
sections shows that $\phi$, $\dot\phi$, $R$ and the square of the shear
$\sigma_{ab}\sigma^{ab}$ tend to zero as $t\to\infty$. Beyond this point
it must be expected that the asymptotics of solutions of different
symmetry classes diverge, as they do in the case without a scalar field.

To the author's knowledge the only classes of homogeneous solutions of the 
Einstein equations coupled to a massive scalar field whose late-time
asymptotics have been analysed further than in this paper are the Bianchi
type I solutions and the isotropic solutions with non-vanishing spatial
curvature. There is a heuristic discussion of the dynamics in these cases
in \cite{belinskii1} and \cite{belinskii2}. One intuitive consideration 
for spacetimes containing only a massive scalar field similar to that 
concerning fluids at the end of Section \ref{matter} follows from 
the equation $dR/dt=-2HR$ which holds in a homogeneous and isotropic 
model with non-zero curvature and the equation 
$d/dt (\sigma_{ab}\sigma^{ab})=-6H(\sigma_{ab}\sigma^{ab})$ which holds
in a Bianchi I spacetime. It suggests that in the latter case the influence 
of the shear at late times should be negligible while in the former case
the curvature should dominate. Indeed the claim in \cite{belinskii1} and
\cite{belinskii2} is that the late-time behaviour for negative curvature 
resembles that of the Milne model. Of course in the case of positive 
curvature the solution might recollapse.

In general it appears that it remains to obtain rigorous results for even
some of the simplest models related to the massive scalar field. It would
be interesting to see how the oscillations of the scalar field interact with
those due to the gravitational field which arise in more complicated 
Bianchi types. For these Bianchi types averaging techniques are necessary 
to analyse the case without scalar field. Cf. \cite{nilsson}, \cite{ringstrom},
\cite{wainwright}. Ideally a unified approach to all these problems should be
developed.

\section{Acknowledgements} 

I thank Spiros Cotsakis and John Miritzis for stimulating discussions.

\end{document}